\documentclass[aps,twocolumn,a4paper]{revtex4}%
\usepackage{psfig}
\usepackage{amssymb}
\usepackage{amsmath}
\usepackage{amsfonts}
\usepackage{graphicx}
\begin{document}

\title{The Casimir effect in the nanoworld}

\author{Cyriaque Genet$^1$, Astrid Lambrecht$^2$, and Serge Reynaud$^2$  }\email{reynaud@spectro.jussieu.fr}
\affiliation{$^1$ Laboratoire des Nanostructures, ISIS, CNRS, ULP, F-67083 Strasbourg Cedex, France}
\affiliation{$^2$ Laboratoire Kastler Brossel, ENS, CNRS, UPMC, Campus Jussieu, F-75252 Paris Cedex 05, France}

\begin{abstract}
The Casimir effect is a force arising in the macroscopic world as a result of 
radiation pressure of vacuum fluctuations. It thus plays a key role in the 
emerging domain of nano-electro-mechanical systems (NEMS).
This role is reviewed in the present paper, with discussions of the influence 
of the material properties of the mirrors, as well as the geometry dependence 
of the Casimir effect between corrugated mirrors. 
In particular, the lateral component of the Casimir force and restoring torque 
between metal plates with misaligned corrugations are evaluated. 
\end{abstract} 

\maketitle

\section{Introduction}
\label{intro}
The Casimir force was predicted in $1948$ by H.B.G. Casimir as an attractive force 
between two perfectly reflecting, plane and parallel mirrors in vacuum \cite{Casimir48}. 
The force has been measured in different experiments with an increasing control of the
experimental conditions \cite{Sparnaay89,Lamoreaux97,Mohideen98,Harris00,%
Ederth00,Bressi02,Decca03prl,ChenPRA04,DeccaAP05,Decca07}.
This has been considered as an important aim which should allow an accurate 
comparison between theoretical predictions and experimental observations 
\cite{Milonni94,LamoreauxResource99,Reynaud01,GenetIAP02,LambrechtPoincare}. 
These advances have been reviewed in a number of papers, for example
\cite{LambrechtPoincare,Bordag01,Milton05} and in a special issue of the 
New Journal of Physics \cite{NJP06}. 

Meanwhile, it has been realized that the Casimir force was a dominant force 
at micron or sub-micron distances, and then clearly an important aspect
in the domain of micro- and nano-oscillators (MEMS, NEMS) 
\cite{BuksPRB2001,ChanScience2001,ChanPRL2001} now emerging from modern 
nanofabrication techniques \cite{EkinciRSI2005}. 
If the Casimir force has been primarly considered as a source of stiction 
between mobile parts, it is now recognized as an essential source of actuation 
to be used in the design of MEMS and NEMS. 

In both fundamental and technological contexts, it is extremely important 
to take into account the real experimental situations which largely
differ from the ideal conditions considered by Casimir. 
We review below some theoretical tools which have shown their efficiency
for a general formulation of the Casimir effect, accounting for the 
material properties of the interacting plates as well as for the effect of 
non planar boundary geometries. 

\section{Idealized Casimir force}
\label{sec:1}
The Casimir force and energy between two perfectly reflecting, plane and parallel mirrors 
immersed in quantum vacuum have the following forms 
\begin{eqnarray}
F_{\rm Cas} = \frac{\pi^{2}\hbar c}{240}\frac{A}{L^{4}}  \ \ , \ \
E_{\rm Cas} = - \frac{\pi^{2}\hbar c}{720}\frac{A}{L^{3}}.    \label{FEcas}
\end{eqnarray}
These expressions correspond to an attractive force $F_{\rm Cas}$ and a binding energy $E_{\rm Cas}$.
Remarquably, they depend only on geometrical quantities, the area $A$ of the mirrors and their 
distance $L$ ($A\gg L^{2}$), and fundamental constants, the Planck constant $\hbar$ and the speed of light $c$. 

\section{Imperfect reflection}
\label{sec:3}
Experiments are performed with real metallic mirrors which good reflectors only at frequencies below 
their plasma frequency $\omega_{\rm P}$ which depends on the properties of the conduction electrons in the metal. 
The effect of imperfect reflection on the Casimir force and energy has been recognized long time ago 
\cite{Lifshitz56,Schwinger78} though it has been described with good accuracy only recently 
\cite{Lamoreaux99,Lambrecht00,KlimPRA00}. We recall below the scattering theory of the Casimir force 
which has been developed and used to this aim \cite{Jaekel91,GenetPRA03,LambrechtNJP06}.

We begin with perfectly plane and parallel mirrors, separated by a distance $L$. 
The two mirrors form a Fabry-Perot cavity and the fluctuations of the intracavity fields 
propagating back and forth along the cavity axis can be calculated in terms of the fluctuations 
of the incoming free-space fields.
The field modes are characterized by their frequency $\omega$, transverse wavevector ${\bf k}$ 
with components $k_{x},k_{y}$ in the plane of the mirrors, and by their polarization $p$. 
Time invariance of the problem, as well as transverse spatial translation invariance 
(along $x$ and $y$) ensure that the frequency, the transverse wavevector and the polarization 
are conserved quantities throughout the scattering process on the cavity. 
The scattering couples only the free vacuum modes with opposite signs for the component 
$k_{z}$ of the wavevector along the longitudinal $z$ axis of the cavity. 
We write $r_{\bf k}^{p}[\omega]$ the reflection amplitude of the mirror $i=1,2$ 
as seen from the inner side of the cavity. 
These amplitudes obey general physical properties of causality, unitarity and high frequency transparency.

The spectral density of the vacuum intracavity fields is changed with respect to that of free-fields outside the cavity. 
The ratio of energy inside the cavity to energy outside the cavity is fixed, for a given mode, by the following function
\begin{eqnarray}
g_{\bf k}^{p}[\omega] = \frac{1-|\rho_{\bf k}^{p}[\omega]|^{2}}{|1-\rho_{\bf k}^{p}[\omega]|^{2}} \ \ , \ \  \rho_{\bf k}^{p}[\omega] = r_{\bf k}^{p}[\omega]_{1}r_{\bf k}^{p}[\omega]_{2}e^{2ik_{z}L}.
\end{eqnarray}
This statement constitues a theorem which has been demonstrated for lossless as well as lossy mirrors \cite{GenetPRA03,Barnett96}. 
It does not depend on the state of the fields and is therefore valid for vacuum fluctuations as well as for thermal fluctuations, 
assuming thermal equilibrium. 
We do not discuss here the issue of thermal dependence of the Casimir effect (see for example the recent review \cite{Brevik06})
and restrict our attention to the zero temperature limit. 

The force is the difference in radiation pressure between inner and outer faces of the mirrors, integrated over all the modes. 
Using analyticity properties, the force and energy may be written as integrals over imaginary frequencies $\omega =i\xi$
\begin{eqnarray}
F&=&\frac{\hbar A}{\pi} \sum_{p}\int\frac{{\rm d}^{2}{\bf k}}{4\pi^{2}}\int_{0}^{\infty}{\rm d}\xi \kappa[i\xi]\frac{\rho_{\bf k}^{p}[i\xi]}{1-\rho_{\bf k}^{p}[i\xi]}   \ \ , \nonumber  \\
E&=&\frac{\hbar A}{2\pi}\sum_{p}\int\frac{{\rm d}^{2}{\bf k}}{4\pi^{2}}\int_{0}^{\infty}{\rm d}\xi \ln \left(1-\rho_{\bf k}^{p}[i\xi]\right).  \label{FEscatt}
\end{eqnarray}
$\kappa[i\xi]=\sqrt{{\bf k}^{2}+\xi^{2} / c^{2}}$ is the longitudinal component of the wavevector evaluated for imaginary frequencies. 

The expressions (\ref{FEscatt}) are regular for any physical model of the reflection amplitudes. 
High frequency transparency of any real mirror ensures that the integrals are convergent, and free 
from the divergences usually associated with the infinitness of vacuum energy. 
They reproduce the Lifshitz expression for the Casimir force \cite{Lifshitz56,Schwinger78} when 
assuming that the metal plates have large optical thickness with reflection amplitudes given by the 
Fresnel laws on the vacuum-bulk interface
\begin{eqnarray}
r_{\bf k}^{\rm TE}[i\xi]&=&-\frac{\sqrt{\xi^{2}\left(\varepsilon[i\xi]-1\right)+c^{2}\kappa^{2}}-c\kappa}{\sqrt{\xi^{2}\left(\varepsilon[i\xi]-1\right)+c^{2}\kappa^{2}}+c\kappa}  \   \ ,  \nonumber  \\
r_{\bf k}^{\rm TM}[i\xi]&=&-\frac{\sqrt{\xi^{2}\left(\varepsilon[i\xi]-1\right)+c^{2}\kappa^{2}}-c\kappa\varepsilon[i\xi]}{\sqrt{\xi^{2}\left(\varepsilon[i\xi]-1\right)+c^{2}\kappa^{2}}+c\kappa\varepsilon[i\xi]}.  \label{Fresnel}
\end{eqnarray}
Here $\varepsilon[i\xi]$ is the dielectric function describing a optical response of the material
inside the mirrors. 
Taken together, relations (\ref{FEscatt}) and (\ref{Fresnel}) reproduce the Lifshitz expression \cite{Lifshitz56}. 
They are known to tend to the original Casimir expression in the limit $\varepsilon \rightarrow \infty$ 
which produces perfectly reflecting mirrors \cite{Schwinger78}. 

We may emphasize at this point that relations (\ref{FEscatt}) are more general than Lifshitz expression
which, incidentally, were not written originally in terms of reflection amplitudes \cite{Katz77}. 
They are valid for example for non-local optical responses of the mirrors provided the reflection
amplitudes are substituted by their possibly more complicated expressions. 
The only limitation, discussed below, is associated with the assumption of specular scattering.

\section{Finite conductivity corrections}
\label{sec:4}
We now review the corrections to the Casimir expression coming from the finite conductivity of the bulk material. 
Here, these corrections are deduced from relations (\ref{FEscatt}), assuming Fresnel laws (\ref{Fresnel}) 
for a local optical response of the bulk material. 
This function may be given by a simple description of the conduction electrons in terms of a plasma model
\begin{eqnarray}
\varepsilon[i\xi]=1+\frac{\omega_{\rm P}^{2}}{\xi^2},
\end{eqnarray}
characterized by a plasma frequency $\omega_{\rm P}$ and wavelength $\lambda_{\rm P}\equiv 2\pi c/\omega_{\rm P}$. 
It may be given by a more realistic representation based upon tabulated optical data 
and which includes the contribution of interband electrons \cite{Lambrecht00}. 

\begin{figure}
\resizebox{0.75\columnwidth}{!}{\includegraphics{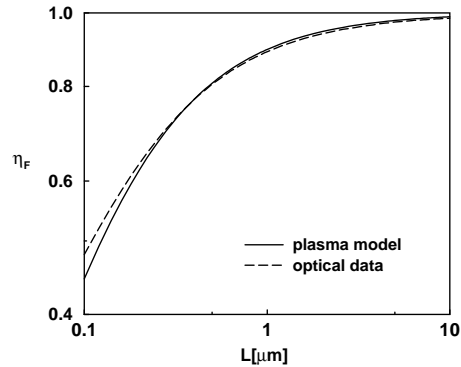} }
\caption{Reduction factor $\eta _{\rm F}$ for the Casimir force between two identical 
Au mirrors at zero temperature as a function of the distance $L$. The solid and dashed 
lines correspond to evaluations based respectively on the plasma model with 
$\lambda_{\rm P}=136$nm and on tabulated optical data \cite{Lambrecht00}.}
\label{fig:1}     
\end{figure}

The corrections to the Casimir effect are conveniently represented in terms of factors measuring the 
reduction of the force and energy with respect to the ideal limit of perfect mirrors
\begin{eqnarray}
F&=&\eta_{\rm F}F_{\rm Cas}  \ \ , \ \   \eta_{\rm F} < 1   \  \ \textrm{and} \nonumber  \\  
E&=&\eta_{\rm E}E_{\rm Cas}  \ \ , \ \   \eta_{\rm E} < 1.
\end{eqnarray}
The results of the calculations are plotted on Fig.(\ref{fig:1}) for Au-covered mirrors.
They are shown as $\eta_{\rm F}$ varying versus the ratio of the cavity length $L$ to 
the plasma wavelength $\lambda_{\rm P}$. 

For metals used in recent experiments, the plasma wavelength lies around $0.1\mu$m ($136$nm for Au and Cu). 
At large distances $L\gg\lambda_{\rm P}$, the ideal Casimir formula is recovered ($\eta_{\rm F}\rightarrow 1$), as expected. 
At short distances, a significant reduction of the force is obtained, with a change in the power law for the variation 
of the force with distance. 
This change can be understood as the result of the Coulomb interaction of surface plasmons at the two vacuum interfaces 
\cite{GenetAFLdB04,Henkel04}. 
This interpretation may be actually generalized to arbitrary distances at the price of a full electromagnetic treatment 
of the plasmonic as well as ordinary photonic modes \cite{Intravaia05,Intravaia07}.
The plasma model is sufficient for a first description of the variation of the force with distance 
but it is not sufficient for a precise comparison. 

First, the relaxation of the conduction electrons has to be accounted for. 
Then, interband transitions are reached for metals like Au, Cu or Al for photon energies of a few eV
and their effect on the optical response has to be taken into account for evaluating the Casimir force 
at short (sub-micron) distances. 
This can be done by using tabulated optical data which are integrated using causality relations \cite{Lambrecht00}.
The result of the corresponding evaluation is shown on Fig.(\ref{fig:1}). 
It is worth stressing that calculations are sensitive to the existing differences in optical data between 
different tabulated sets \cite{Pirozhenko06}. 
This means that an imperfect knowledge of the optical properties of the mirrors used in the experiment 
is a source of uncertainty in the experiment-theory comparison. 
Ideally, if the aim is to have a reliable theoretical evaluation of the Casimir force to be compared
with experiments, it is necessary to measure the reflection amplitudes \textit{in situ}.

\section{Silicon slab mirrors}
\label{sec:5}
As stressed in the introduction, the relevance of the Casimir effect on nanosystems 
calls for a precise understanding not only of the influence of material optical properties 
on the Casimir force, but also of the influence of geometrical parameters, 
such as the thickness of the coatings \cite{IannuzziPNAS2004,LisantiPNAS2005} 
or the thickness of the mirrors themselves. 
In this context, structures made of silicon, the reference material used in nano-fabrication processes, 
are particularly interesting to study  \cite{LambrechtEPL2007,Chen07}.

\begin{figure}
\resizebox{0.75\columnwidth}{!}{\includegraphics{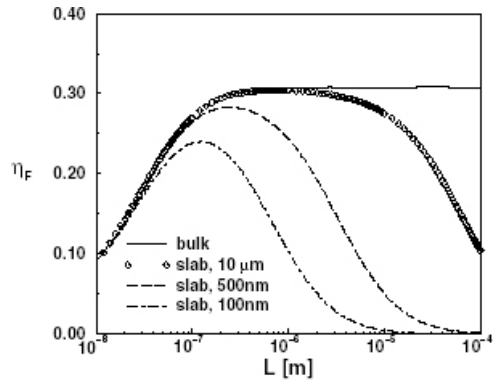} }
\caption{Reduction factor and absolute value of the Casimir force per unit surface (bottom) between two silicon slabs as a function of slab separation for different slab thicknesses as well as for a silicon bulk mirror for comparison.}
\label{fig:2}     
\end{figure}

The reflection amplitude corresponding to a slab of finite thickness $D$ 
is different from the bulk expression and is given through a Fabry-Perot formula
\begin{eqnarray}
r_{\bf k}^{p}[i\xi]_{\rm slab}&=&r_{\bf k}^{p}[i\xi]\frac{1-e^{-2\delta}}{1-(r_{\bf k}^{p}[i\xi])^{2}e^{-2\delta}}  \ \ , \nonumber  \\ 
\delta &=&\frac{D}{c}\sqrt{\xi^{2}(\varepsilon[i\xi]-1)+c^{2}\kappa^{2}}.   \label{Rslab}
\end{eqnarray}
$r_{\bf k}^{p}[i\xi]$ is the bulk reflection amplitude given by (\ref{Fresnel}). 
Using these reflection amplitudes for calculating the Casimir force between two Si slabs, 
interesting behaviours have been noted \cite{LambrechtEPL2007} which differ from the situation of metallic mirrors. 
In particular, it was shown that the material thickness has a stronger influence on the Casimir force for Si slabs than for Au slabs. 
For Si, the force decreases as soon as the slab separation $L$ is larger than the slab thickness $D$, as seen on Fig.(\ref{fig:2}). 
In contrast to metals which become perfect reflectors in the limit of zero frequency, 
Si is a semiconductor with a finite transverse plasma frequency $\omega_{0}$ 
corresponding to a cut-off wavelength $\lambda_{0}=2\pi c / \omega_{0}\sim 286$nm. 
For cavity length $L$ smaller than this cut-off wavelength, Si tends to become transparent. 
The associated optical thickness $\delta$ given in Eq.(\ref{Rslab}) is large, 
so that the Si slab behaves like a bulk Si mirror with low reflectivity at high frequency. 
The Casimir force is then much smaller than the perfect reflection limit of Eq.(\ref{FEcas}). 
On the other hand, at low frequencies $\omega\ll\omega_{0}$, one will have $\delta\ll 1$ together 
with $c\kappa\rightarrow 0$, low frequencies being predominant at large distances. 
In this latter case, the slab is transparent again, and the Casimir force between two Si slabs 
is decreased when $L\geq D$. 
This result can have interesting consequences for nanostructures as it opens a way to control 
the magnitude of the Casimir force and possibly eliminate an unwanted Casimir source of stiction.
From a fundamental point of view, it also offers a new solution to study the comparison between
experiment and theory of the Casimir force \cite{Chen07}

\section{Geometry and the Casimir effect}
\label{sec:6}
Geometry effects are expected to lead to a rich variety of behaviours in the Casimir physics 
\cite{Balian7778,Plunien86,Balian0304}.
Recent advances make it possible to explore this interplay, both from experimental and theoretical point of views. 
This also offers new possibilities for tailoring the Casimir force through specific designs 
\cite{EmigEPL03}. 

Force and energy evaluations between non planar mirrors are commonly obtained using the so-called 
proximity-force approximation (PFA) \cite{Derjaguin68,Langbein71}. 
This approximation amounts to an averaging of plane-plane contributions over the distribution of 
local interplate separations defined by the chosen geometry. For the energy, the PFA leads to
\begin{eqnarray}
E_{\rm PFA} = \int\frac{{\rm d}^{2}{\bf r}}{A}E_{\rm PP}\left(\ell)\right)  \ , \ 
\ell \equiv L-h_{1}({\bf r})-h_{2}({\bf r}), \label{PFArough}
\end{eqnarray}
with $h_{1}({\bf r})$ and $h_{2}({\bf r})$ the surface profiles of each mirrors.
Such profiles can be described by their spectra evaluated over the surface $A$ of the mirrors
\begin{eqnarray}
\int \frac{{\rm d}^{2} {\bf r}}{A}h_{i}({\bf r})h_{j}({\bf r})=\int\frac{{\rm d}^{2}{\bf k}}{(2\pi)^{2}}h_{i}[{\bf k}]h_{j}[-{\bf k}] \ , \ i,j=1,2   
\end{eqnarray}
with $h_{i}[{\bf k}]$ the Fourier transform of $h_{i}({\bf r})$, and by the associated correlation lengths $\ell_{\rm C}$. 
When they are smaller than the other length scales, the amplitudes of deformations can be considered as perturbations. 
A second order expansion in the profiles can thus be performed leading to
\begin{eqnarray}
E_{\rm PFA} = E_{\rm PP}+\frac{1}{2}\frac{\partial^{2}E_{\rm PP}}{\partial L^{2}}\int \frac{{\rm d}^{2} {\bf r}}{A}(h_{1}({\bf r})+h_{2}({\bf r}))^{2}.  \label{PFAgen}
\end{eqnarray}
The trivial first-order term has been discarded, assuming that the deformations have zero spatial averages 
$\int {\rm d}^{2}{\bf r}h_{i=1,2}({\bf r}) / A =0$.  

The evaluation of the effect of geometry through the PFA, based on a summation procedure over local contributions 
assumes some additivity property of the Casimir effect, whereas the Casimir force is known not to be additive. 
The PFA can only be accurate for surfaces which can be considered as nearly plane with respect to other scales 
such as the separation distance $L$ \cite{GenetEPL03}. 
For example, it allows one to calculate the Casimir force in the plane-sphere (PS) configuration as
\begin{eqnarray}
F_{\rm PS}=\frac{2\pi R}{A}E_{\rm PP},  \ \ \textrm{with} \ \ L\ll R, \label{PFAPS}
\end{eqnarray}
where $E_{\rm PP}$ is the Casimir energy in the plane-plane (PP) geometry.
Most recent experiments are performed in the plane-sphere geometry which is much simpler to control 
than the plane-plane configuration. 
The PFA is here expected to be valid provided the radius $R$ of the sphere is much larger 
than the distance $L$ of closest approach. 

But the PFA certainly fails for describing more general surface profiles. 
As far as plate deformations are concerned, it can only be valid in the limit $\ell_{\rm C}\gg L$ 
which corresponds to a trivial specular description of the reflection process on the surfaces \cite{MaiaNetoPRA05}. 
For the general case, a description of non specular scattering process on mirrors is available
for analyzing the connection between geometry and the Casimir effect \cite{MaiaNetoPRA05}. 
An expression for the Casimir energy between parallel mirrors with arbitrary surface profiles 
has been derived in \cite{LambrechtNJP06,RodriguesPRA2007} 
\begin{eqnarray}
E=\hbar\int\limits_{0}^{\infty}\frac{{\rm d}\xi}{2\pi}{\rm Tr} \ {\rm ln}\left(1-\rm{R}_{1}\left(i\xi\right)e^{-K\left(i\xi\right)L}\rm{R}_{2}\left(i\xi\right)e^{-K\left(i\xi\right)L}\right)\  \  \label{formule}
\end{eqnarray}
This expression is based on non-specular reflection matrices $\rm{R}_{1}$ and $\rm{R}_{2}$ associated to each mirror. 
While the operator $e^{-K\left(i\xi\right)L}$ corresponds to propagation of the field between the two mirrors, 
and is diagonal in the plane-wave basis with elements given by $K(i\xi) =\sqrt{{\bf k}^2+\xi^2 / c^{2}}$, 
the two matrices $\rm{R}_{1}$ and $\rm{R}_{2}$ are non-diagonal on plane-waves. 
This corresponds to a mixing of polarizations and wavevectors, due to non-specularity diffraction on the gratings
formed by the profiles on the surfaces of the mirrors. 

As it is reviewed below, this formula (\ref{formule}) has been used to evaluate the effect of surface roughness 
\cite{MaiaNetoPRA05} or corrugations on the Casimir force \cite{RodriguesEPL2006,RodriguesPRL2006}. 
Analytical expressions have been derived through a perturbative treatment, with the roughness or the corrugation 
amplitudes taken as the smallest length scales involved in the problem. 
The effect of the optical response of the metal has been included in these calculations. 
It is worth stressing that this formula has a wider range of validity. 
It can in principle describe structured plates with large corrugation amplitudes,
as well as material properties not limited to a simple plasma model.
The only task for a quantitative evaluation of the Casimir force or energy 
is to obtain the actual form of the reflection operators $\rm{R}_{1}$ and $\rm{R}_{2}$ 
to be inserted into Eq.(\ref{formule}).

\section{Roughness correction}
\label{sec:7}
A correction to the Casimir force that must be accounted for is the effect of surface roughness, 
intrinsic to any real mirror. 
This effect is analyzed in recent experiments through procedures based on the PFA \cite{BordagPLA95,KlimPRA99}. 
The general formula (\ref{formule}) has been used to go beyond this approximation \cite{MaiaNetoPRA05}. 
As already stressed, the roughness amplitude must be the smallest length scale for perturbation theory to hold. 
Meanwhile, the plasma wavelength, the mirror separation and the roughness correlation length 
may have arbitrary relative values with respect to each other. 

\begin{figure}
\resizebox{0.75\columnwidth}{!}{\includegraphics{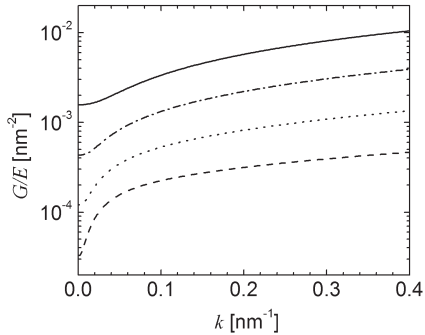} }
\caption{Variation of $G/E_{\rm PP}$
versus $k$ 
for the distances $L=50$nm (solid line), 
$L=100$nm (dashed-dotted line), $L=200$nm (dotted line), 
and 
$L=400$nm (dashed line). We take $\lambda_{\rm P}=136$nm.}
\label{fig:3}     
\end{figure}

We remind that the roughness profiles are defined with respect to reference mirror planes separated by the mean distance $L$. 
We assume that profiles have zero averages and show no cross-correlations. 
We also suppose that the area $A$ of each plate is large enough to include many correlation areas ($A\gg \ell_{\rm C}^{2}$), 
so that surface averages are identical to statistical averages. 
Up to second order in the profiles, the correction to the Casimir energy may thus be written as follows
\begin{eqnarray}
\delta E_{\rm PP}=\int\frac{{\rm d}^{2}{\bf k}}{(2\pi)^{2}}G_{\rm rough}[{\bf k}]\sigma[{\bf k}].  \label{beyondPFArug}
\end{eqnarray}
Here $\sigma[{\bf k}]$ corresponding to the roughness spectrum added over the two plates. 
$G_{\rm rough}[{\bf k}]$ is a spectral sensitivity to roughness of the Casimir energy. 
Due to cylindrical symmetry with respect to rotations in the transverse plane, it only depends on $k=|{\bf k}|$. 
This dependence reveals that the roughness correction does not only depend on the root-mean-square (rms) roughness, 
but also on the spectral distribution of the roughness. 
Fig.(\ref{fig:3}) displays $G_{\rm rough}[k]$ normalized by $E_{\rm PP}$ as it has been calculated 
for Au-covered mirrors and for various interplate distances. 
The rich behaviours of $G_{\rm rough}[k]$ as a function of the length scales is discussed in \cite{MaiaNetoEPL05}.

What we want to stress here is that this function describes deviations from the PFA.
The width of the roughness spectrum $\sigma[{\bf k}]$ is indeed fixed by the inverse of the correlation length $\ell_{\rm C}$. 
When this spectrum is contained in the region where $G_{\rm rough}[k]$ remains close to its secular limit $G_{\rm rough}[0]$, 
we can approximate Eq.(\ref{beyondPFArug}) as proportional to the rms roughness
\begin{eqnarray}
\delta E_{\rm PP}\simeq G_{\rm rough}[0]\langle h_{1}^{2}+h_{2}^{2}\rangle.  \label{PFArug}
\end{eqnarray}
This corresponds effectively to the PFA expression, as the consequence of a theorem
which was proved in \cite{MaiaNetoPRA05} 
\begin{eqnarray}
G_{\rm rough}[k\rightarrow 0]= \frac{1}{2}\frac{\partial^{2}E_{\rm PP}}{\partial L^{2}}.
\label{PFtheorem}
\end{eqnarray}
Equation (\ref{PFtheorem}) is nothing but a properly stated ``Proximity Force Theorem''.
It can however not be confused with the ``Proximity Force Approximation'' (\ref{PFArug}) 
which is a good approximation only for smooth enough mirrors, that is also for 
large enough roughness correlation lengths $\ell_{\rm C}$.

In the general case, the PFA result (\ref{PFArug}) underestimates the effect of roughness. 
When performing the theory-comparison, one has therefore to carefully assess the 
roughness correction by measuring the roughness spectra \textit{in situ} and
using the roughness sensitivity function as given in \cite{MaiaNetoPRA05,MaiaNetoEPL05}.
The PFA can only be used if $\ell_{\rm C}$ has been proven to be large enough or,
in a looser way, when the roughness correction has been estimated to have a negligible
value.

\section{Lateral force between corrugated plates}
\label{sec:8}
As the roughness effect remains a small correction to the Casimir force,
it seems difficult to measure the deviation from PFA regime and check
its agreement with theory. 
Fortunately, there exists an experimental configuration showing more promising perspectives
as a potential probe of the non-trivial interplay between the Casimir effect and geometry.

This configuration corresponds to periodically corrugated metallic plates placed face to face
in vacuum, so that a lateral component of the Casimir force arises due to the breaking of 
the transverse translational invariance \cite{Golestanian}. 
A recent experiment has demonstrated the feasibility of a lateral force measurement at 
separation distances of the order of $\sim 100$nm \cite{ChenPRL02}. 
Since it would disappear in the absence of corrugation, the lateral force should not be
considered as a small correction to the otherwise dominant normal Casimir force, as it was 
the case for the study of roughness. 
As we will see below, the deviation from PFA indeed appears as a factor in front of
the lateral force, so that a precise measurement of this force would test
in a crucial manner the interplay between Casimir effect and geometry \cite{RodriguesPRL2006}.
As the experiments are performed at short distances, it cannot be described with the assumption
of perfect reflection, where analytical results are available \cite{EmigPRA03,EmigPRL05}. 
Again, the general scattering formula (\ref{formule}) shows the ability to give an estimation 
for the lateral force for arbitrary relative values of the length scales $\lambda_{\rm C}$, $\lambda_{\rm P}$ and $L$, 
provided the corrugation amplitudes $a_{i=1,2}$ remain the smallest length scales of the problem.

\begin{figure}
\resizebox{0.45\columnwidth}{!}{\includegraphics{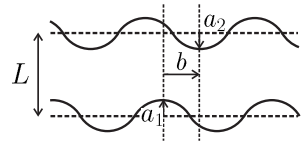} }
\caption{Surface profiles considered for the lateral component of
the Casimir force. Both surfaces have a sinusoidal corrugation with
$a_1$ and $a_2$ being the corrugation amplitudes, $b$ the mismatch
between the two sinusoidal functions.}
\label{fig:4}     
\end{figure}

We consider two metallic mirrors, both sinusoidally corrugated along one dimension, 
with the same corrugation wavelength $\lambda_{C}$, separated by a distance $L$ and 
facing each other with a relative spatial mismatch $b$ between the corrugation crests -see Fig.(\ref{fig:4}). 
The profiles $h_{i=1,2}({\bf r})$, ${\bf r}=(x,y)$, of the two uniaxial (along $y$) 
corrugated mirrors are defined by the two functions $h_{1} = a_{1}\cos\left(k_{\rm C}x\right)$ 
and $h_{2}=a_{2}\cos\left(k_{\rm C}\left(x-b\right)\right)$
with $k_{\rm C}= 2\pi / \lambda_{\rm C}$ the wavevector associated to the corrugation wavelength $\lambda_{\rm C}$. 
We take both profiles with zero spatial averages. 
At the second order in the corrugations, cross-terms of the form $a_{1}a_{2}$ appear which
contribute to the lateral force because the energy depends on the transverse mismatch $b$. 

This fact, a consequence of the correlation between the two corrugation profiles, induces a 
contrast with the case of roughness where the effect was associated with quadratic terms $h_{i=1,2}^{2}$. 
It implies that the evaluation of the lateral force only involves first-order non-specular amplitudes 
calculated on each mirror separately. 
The full calculation gives the second-order correction to the Casimir energy induced by the corrugations 
\begin{eqnarray}
\delta E_{\rm PP}=A\frac{a_{1}a_{2}}{2}\cos (k b) G_{\rm C}[k].
\label{secorder}
\end{eqnarray}
The function $G_{\rm C}[{\bf k}]$ is given in \cite{RodriguesPRL2006} and does only depend on the modulus $k$ of ${\bf k}$. 
Here again, the PFA regime is recovered in the $k\rightarrow 0$ limit, as a consequence 
\begin{eqnarray}
G_{\rm C}[k\rightarrow 0]=\frac{1}{2} \frac{\partial^{2}E_{\rm PP}}{\partial L^{2}}.
\end{eqnarray}
This theorem is ensured, for any specific model of the material medium, by the fact that $G_{\rm C}$ 
is given for $k\rightarrow 0$ by the specular limit of non-specular reflection amplitudes \cite{MaiaNetoPRA05}.   

In order to compare with experiments, we consider the expression of the lateral force in the plane-sphere configuration. 
It is derived from the plane-plane configuration using the PFA, reliable as long as $L\ll R$. 
In fact, there is no interplay between curvature and corrugation provided $RL\gg\lambda_{\rm C}^{2}$, 
a condition met in the experiment reported in \cite{ChenPRL02}. 
\begin{figure}
\resizebox{0.85\columnwidth}{!}{\includegraphics{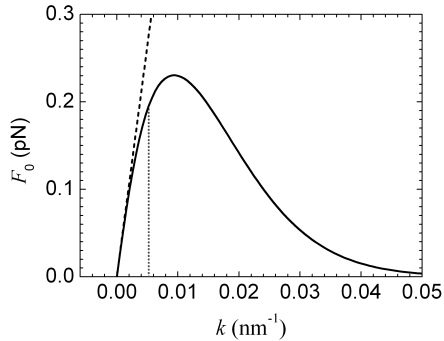} }
\caption{Lateral force amplitude for the plane-sphere setup, as a function of $k,$ 
with figures taken from \cite{ChenPRL02}. The experimental value $k=0.0052$nm$^{-1}$ is indicated by the vertical dashed line.}
\label{fig:5}     
\end{figure}
From Eq.(\ref{PFAPS}), the lateral force in the plane-sphere geometry is eventually given as \cite{RodriguesPRL2006}
\begin{eqnarray}
F_{\rm PS}^{\rm lat}=-\frac{\partial }{\partial b}E_{\rm PS}^{\rm lat}=\pi a_{1}a_{2}kR\sin (kb )
\int_{\infty}^{L}{\rm d}L^{\prime}G[k,L^{\prime}].
\end{eqnarray}
The force is plotted in Fig.(\ref{fig:5}) as a function of $k$, with length scales 
$\lambda_{\rm C}$, $\lambda_{\rm P}$ and $L$ fitting the experimental values \cite{ChenPRL02}. 
As the corrugation amplitudes are not small enough in the experiment to meet the perturbation
condition, the theory and experiment can unfortunately not be compared directly.
The plot on Fig.(\ref{fig:5}) nevertheless shows the interesting fact that the length scales 
taken from the experiment, with $k$ indicated by the vertical dashed line, 
clearly fall outside the PFA sector in the perturbative calculation. 
For related implications, we refer the reader to the discussions in \cite{RodriguesPRA2007}.  
 
It appears clearly on the figure that the PFA overestimates the magnitude of the lateral force 
for arbitrary $k$. 
We also note that the PFA prediction for the force scales as $k$ when $k$ increases from zero.
At larger values of $k$ in contrast, the lateral force decreases. 
This is due to the one-way propagation factor separating the two first-order non-specular reflections 
at each plate, given as a decresing exponential $e^{-kL}$ in the high $k$ limit \cite{RodriguesPRL2006}. 
It follows that there is a maximal force when $k$ is varied.
It corresponds to $k=9\times 10^{-3}$nm$^{-1}$ with the other length scales corresponding to the experiment.
The ratio $L / \lambda_{\rm C} = 1 / \pi$ is thus falling outside the PFA sector which confirms 
that a lateral force measurement is an excellent candidate for probing deviations from the PFA.

\section{Torque}
\label{sec:9}
Another interesting observable for exploring the non-trivial geometry dependence of the Casimir energy 
is the torque arising when the corrugations of the two plates are misaligned. 
With this angular mismatch between the corrugations, rotational symmetry is broken and induces 
a restoring torque between the plates.

The calculations are quite similar to those which were done for aligned corrugated surfaces, 
in particular because the same non-specular reflection coefficients are used to describe each plate. 
The second-order correction is still given by the sensitivity function $G_{\rm C}[{\bf k}]$ 
which does only depend on the modulus of the corrugation wavevector ${\bf k}$. 
The difference with the lateral force case lies only in the fact that the corrugation profiles 
$h_{i=1,2}({\bf r})=a_{i}\cos ({\bf k}_{i}\cdot {\bf r} - kb_{i})$ corresponds to different
corrugation wavevectors ${\bf k}_{i=1,2}$ having however the same modulus $k=2\pi /\lambda_{\rm C}$. 
The angular mismatch between ${\bf k}_{1}$ and ${\bf k}_{2}$ is given by the angle $\theta$. 
The parameters $b_{i}$ represent lateral displacements with respect to the configuration 
with a line of maximum height at the origin. 
We assume that the corrugation $h_{2}$ is restricted to a rectangular section of area $L_{x}L_{y}$ 
centered at $x=b_{2},y=0$ and much larger than $L^{2}$ so that diffraction at the borders can be neglected. 
With these assumptions, and in the limit of long corrugation lines $kL_{y}\gg 1$ with $L_{x}$ smaller or of the order of $L_{y}$, 
the energy correction per unit area is given in \cite{RodriguesEPL2006} as
\begin{eqnarray}
\frac{\delta E_{\rm PP}}{L_{x}L_{y}}=\frac{a_{1}a_{2}}{2}G_{\rm C}[k]\cos (kb)\frac{\sin (kL_{y}\theta /2)}{kL_{y}\theta /2}.  
\label{torque}
\end{eqnarray}
The spatial coefficient $b=b_{2}\cos \theta -b_{1}$ is the relative lateral displacement along the direction ${\bf k}_{1}$. 
As expected by symmetry, this correction is invariant under the transformation $\theta \rightarrow - \theta$ 
and $\theta\rightarrow \pi -\theta$ due to the fact that the corrugation lines have no orientation. 
The case $\theta =0$ corresponds to the result of pure lateral displacement discussed in the preceding section.

\begin{figure}
\resizebox{0.75\columnwidth}{!}{\includegraphics{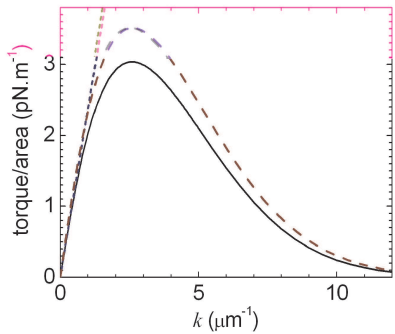} }
\caption{Maximum torque per unit area as a function of $k=2\pi / \lambda_C$ with $L=1\mu$m. Additional parameters
are chosen as in section \ref{sec:8}. The solid line corresponds to the theory presented
in \cite{RodriguesEPL2006}. We also plot the results for perfect reflectors (dashed
line) and PFA with plasma (Au) model (dotted line). }
\label{fig:6}     
\end{figure}

The scale of the energy variation with $b$ and $\theta$ is set by the parameter $ \lambda_{\rm C} / L_{y}$. 
In fact, if plate $2$ is released after a rotation of $\theta >\lambda_{\rm C} / L_{y}$, 
its subsequent motion is a combination of rotation and lateral displacements. 
Rotation is favored over lateral displacements for $\theta <\lambda_{\rm C} / L_{y}$ 
(see Fig.(1) in \cite{RodriguesEPL2006}). 
The torque $\tau = - \partial \delta E_{\rm PP} / \partial\theta $ is evaluated 
in \cite{RodriguesEPL2006} for corrugated Au mirrors, with corrugation amplitudes $a_{1}=a_{2}=14$nm, 
corrugation length $L_{y}=24\mu$m and separated by a distance of $L=1\mu$m. 
It is maximum at $\theta = 0.66 \lambda_{\rm C} / L$ and is plotted in Fig.(\ref{fig:6}) as a function of $k$. 
It starts increasing linearly with $k$ in the $k\rightarrow 0$ PFA sector and for the same reason 
as the lateral force, it decreases exponentially in the high-$k$ limit.
As is clear on Fig.(\ref{fig:6}), the PFA overestimates the magnitude of the torque by a factor of the order 
of $2$ at the peak value of the torque. 
The discrepancy even increases with $k$, since smaller values of $k$ correspond to smoother surfaces. 
The conditions are gathered up towards a direct experimental evidence of a non-trivial effect of geometry. 

Fig.(\ref{fig:6}) also displays the torque when evaluated between perfect metallic corrugated mirrors \cite{EmigPRA03}. 
The corresponding deviation with respect to the calculation given by Eq.(\ref{torque}) stresses that at a separation 
distance of $L=1\mu$m, the optical response of the metal must be accounted for in an accurate evaluation of the torque. 
The perfect conductor limit is reached only if the plasma wavelength 
$\lambda_{\rm P}$ is the smallest length scales (apart from the corrugation amplitudes) of the problem. 

\section{Conclusion}
New perspectives for studying the interplay between Casimir effect and geometry are today clearly visible. 
The theoretical formalism is better and better mastered, so that a rich variety of configurations can be studied.
Meanwhile, novel experimental capabilities are available, allowing one to address challenging questions. 
Proposals have been recently made for measuring the torque between birefringent dielectric disks \cite{MundayPRA2005}. 
A measurement between metallic corrugated mirrors seems to be more easily accessible, with the torque turning out 
to be up to three orders of magnitude higher than the torque between dielectric plates, for comparable separation distance. 
At the same time, alternative routes are explored in order to probe quantum vacuum geometrical effects \cite{RodriguezPRL2007}. 
Cold atoms techniques also look like particularly promising, as they should allow one to see deviations from the PFA 
on the lateral component of the Casimir-Polder force, with a Bose-Einstein condensate used as a local probe 
trapped close to a corrugated surface \cite{DalvitPRL08}. 
These trends suggest that demonstrations of non-trivial effects of geometry should be within reach.

\newcommand{\REVIEW}[4]{\textrm{#1} \textbf{#2}, #3, (#4)}
\newcommand{\Review}[1]{\textrm{#1}}
\newcommand{\Volume}[1]{\textbf{#1}}
\newcommand{\Book}[1]{\textit{#1}}
\newcommand{\Eprint}[1]{\textsf{#1}}
\def\etal{\textit{et al}}

{}


\begin{thebibliography}{99}

\bibitem{Casimir48} H.B.G. Casimir, Proc. K. Ned. Akad. Wet.
\textbf{51}, 793 (1948)

\bibitem{Sparnaay89}  M.J. Sparnaay, in \textit{Physics in the Making} eds. A. Sarlemijn and M.J. Sparnaay (North-Holland, 1989) p. 235
and references therein

\bibitem{Lamoreaux97} S.K.L. Lamoreaux, Phys. Rev. Lett. \textbf{78}, 5 (1997) 

\bibitem{Mohideen98} U. Mohideen and A. Roy, Phys. Rev. Lett. \textbf{81}, 4549 (1998) 

\bibitem{Harris00} B.W. Harris, F. Chen, and U. Mohideen, Phys. Rev.
\textbf{A62}, 052109 (2000)

\bibitem{Ederth00} Th. Ederth, Phys. Rev. \textbf{A62}, 062104 (2000) 

\bibitem{Bressi02} G. Bressi, G. Carugno, R. Onofrio, and G. Ruoso, Phys.
Rev. Lett. \textbf{88}, 041804 (2002) 

\bibitem{Decca03prl} R.S. Decca, D. L\'opez, E. Fischbach, and D.E. Krause,
Phys. Rev. Lett. \textbf{91}, 050402 (2003) and references
therein

\bibitem{ChenPRA04}  F. Chen, G.L. Klimchitskaya, U. Mohideen, and V.M. Mostepanenko, 
Phys. Rev. \textbf{A69}, 022117 (2004)  
 
\bibitem{DeccaAP05} R.S. Decca, D. L\'{o}pez, E. Fischbach, G.L. Klimchitskaya, D.E.
Krause, and V.M. Mostepanenko, Annals Phys. \textbf{ 318}, 37 (2005)

\bibitem{Decca07} R.S. Decca, D. L\'{o}pez, E. Fischbach, G.L. Klimchitskaya, D.E.
Krause, and V.M. Mostepanenko, Phys. Rev. \textbf{D75}, 077101 (2007)  

\bibitem{Milonni94}  P.W. Milonni, \textit{The quantum vacuum} (Academic, 1994)

\bibitem{LamoreauxResource99} S.K. Lamoreaux, Resource Letter in
Am. J. Phys. \textbf{67}, 850 (1999) 

\bibitem{Reynaud01}  S. Reynaud, A. Lambrecht, C. Genet and M.T. Jaekel, C. R. Acad. Sci. Paris \textbf{IV-2}, 1287 (2001) [\Eprint{arXiv:quant-ph/0105053}]

\bibitem{GenetIAP02}  C. Genet, A. Lambrecht and S. Reynaud, in
\textit{On the Nature of Dark Energy} eds. U. Brax, J. Martin, J.P. Uzan, 121
(Frontier Group, 2002) [\Eprint{arXiv:quant-ph/0210173}]

\bibitem{LambrechtPoincare}  A. Lambrecht and S. Reynaud,
Poincar\'e Seminar on Vacuum Energy and Renormalization
\textbf{1}, 107 (2002) [\Eprint{arXiv:quant-ph/0302073}]

\bibitem{Bordag01}  M. Bordag, U. Mohideen, and V.M. Mostepanenko,
Phys. Reports \textbf{353}, 1 (2001) and references therein

\bibitem{Milton05} K.A. Milton, 
J. Phys. \textbf{A20}, 4628 (2005)  

\bibitem{NJP06}  \textit{Focus on Casimir Forces}, 
 eds. R. Barrera and S. Reynaud, New J. Phys. \textbf{8} (2006)
http://www.iop.org/EJ/abstract/1367-2630/8/10/E05.

\bibitem{BuksPRB2001} E. Buks and M.L. Roukes, Phys. Rev. \textbf{B63}, (2001) 033402 

\bibitem{ChanScience2001} H.B. Chan, V.A. Aksyuk, R.N. Kleinman, D.J. Bishop, and F. Capasso, Science \textbf{291}, 1941 (2001)  

\bibitem{ChanPRL2001} H.B. Chan, V.A. Aksyuk, R.N. Kleinman, D.J. Bishop, and F. Capasso, Phys. Rev. Lett. \textbf{87}, 211801 (2001)  

\bibitem{EkinciRSI2005} K.L. Ekinci and M.L. Roukes, Rev. Sci. Instrum. \textbf{76}, 061101 (2005)  


\bibitem{Lifshitz56}E.M Lifshitz, Sov. Phys. JETP \textbf{2}, 73 (1956)  

\bibitem{Schwinger78} J. Schwinger, L.L. de Raad, and K.A. Milton, Ann. Phys.
\textbf{115}, 1 (1978)  

\bibitem{Lamoreaux99} S.K. Lamoreaux, Phys. Rev. \textbf{A59}, R3149 (1999)  

\bibitem{Lambrecht00} A. Lambrecht and S. Reynaud, Euro. Phys. J.
\textbf{D8}, 309 (2000) 

\bibitem{KlimPRA00} G.L. Klimchitskaya, U. Mohideen, and V.M. Mostepanenko,
Phys. Rev. \textbf{A61}, 062107 (2000) 

\bibitem{Jaekel91} M.T. Jaekel and S. Reynaud, J. Physique
\textbf{I-1}, 1395 (1991) [\Eprint{arXiv:quant-ph/0101067}]

\bibitem{GenetPRA03} C. Genet, A. Lambrecht, and S. Reynaud, Phys. Rev.
\textbf{A67}, 043811 (2003)  

\bibitem{LambrechtNJP06}  A. Lambrecht, P.A. Maia Neto and S. Reynaud, New J. Phys. \textbf{8}, 243 (2006)  

\bibitem{Barnett96}  S.M. Barnett, C.R. Gilson, B. Huttner, and N. Imoto,
Phys. Rev. Lett. \textbf{77}, 1739 (1996)  

\bibitem{Brevik06} I. Brevik, S.A. Ellingsen, and K. Milton, New J. Phys. \textbf{8}, 236 (2006) 

\bibitem{Katz77}  E.I. Kats, Sov. Phys. JETP \textbf{46}, 109 (1977)  


\bibitem{GenetAFLdB04} C. Genet, F. Intravaia, A. Lambrecht, and S. Reynaud,
Ann. Found. L. de Broglie \textbf{29}, 311 (2004)  
[\Eprint{arXiv:quant-ph/0302072}]

\bibitem{Henkel04} C. Henkel, K. Joulain, J.Ph. Mulet, and J.J. Greffet,
Phys. Rev. \textbf{A69}, 023808 (2004)

\bibitem{Intravaia05}  F. Intravaia and A. Lambrecht, Phys. Rev. Lett. \textbf{94}, 110404 (2005)  

\bibitem{Intravaia07}  F. Intravaia, C. Henkel, and A. Lambrecht, Phys. Rev. \textbf{A76}, 033820 (2007) 

\bibitem{Pirozhenko06} I. Pirozhenko, A. Lambrecht, and V.B. Svetovoy,
New J. Phys. \textbf{8}, 238 (2006) 


\bibitem{IannuzziPNAS2004} D. Iannuzzi, M. Lisanti, and F. Capasso, Proc. Nat. Ac. Sci. USA \textbf{101}, 4019 (2004)  

\bibitem{LisantiPNAS2005} M. Lisanti, D. Iannuzzi, and F. Capasso, Proc. Nat. Ac. Sci. USA \textbf{102}, 11989 (2005)  

\bibitem{LambrechtEPL2007} A. Lambrecht, I. Pirozhenko, L. Duraffourg, and P. Andreucci, Europhys. Lett. \textbf{77}, 44006 (2007)
 

\bibitem{Chen07} F. Chen, G.L. Klimchitskaya, V.M. Mostepanenko and 
U. Mohideen, Phys. Rev. \textbf{B76}, 035338 (2007) 


\bibitem{Balian7778} R. Balian and B. Duplantier, Ann. Phys. NY \textbf{104}, 300 (1977); 
\textbf{112}, 165 (1978) 

\bibitem{Plunien86}G. Plunien, B. Muller and W. Greiner, Phys. Reports \textbf{134}, 87 (1986) 

\bibitem{Balian0304} R. Balian, 
in \textit{Poincar\'e Seminar 2002 `Vacuum Energy'}, 
eds. B. Duplantier and V. Rivasseau 
(Birkhäuser, 2003), p. 71;
R. Balian and B. Duplantier, in
\textit{15th SIGRAV Conference on General Relativity and Gravitation}, 
[\Eprint{arXiv:quant-ph/0408124}]

\bibitem{EmigEPL03} T. Emig, Europhys.Lett. \textbf{62}, 466 (2003)  


\bibitem{Derjaguin68}  B.V. Deriagin, I.I. Abrikosova, and E.M. Lifshitz,
Quart. Rev. \textbf{10}, 295 (1968)   

\bibitem{Langbein71}
D. Langbein, J. Phys. Chem. Solids \textbf{32}, 1657 (1971)  

\bibitem{GenetEPL03} C. Genet, A. Lambrecht, P.A. Maia Neto, and S. Reynaud,
Europhys. Lett. \textbf{62}, 484 (2003)  


\bibitem{MaiaNetoPRA05} P.A. Maia Neto, A. Lambrecht, and S. Reynaud,
Phys. Rev. \textbf{A72}, 012115 (2005)  

\bibitem{RodriguesPRA2007} R.B. Rodrigues, P.A. Maia Neto, A. Lambrecht, and S. Reynaud, 
Phys. Rev. \textbf{A75}, 062108 (2007) 

\bibitem{RodriguesEPL2006} R.B. Rodrigues, P.A. Maia Neto, A. Lambrecht, and S. Reynaud, 
Europhys. Lett. \textbf{76}, 822 (2006)  

\bibitem{RodriguesPRL2006} R.B. Rodrigues, P.A. Maia Neto, A. Lambrecht, and S. Reynaud, 
Phys. Rev. Lett. \textbf{96}, 100402 (2006)  


\bibitem{BordagPLA95} M. Bordag, G.L. Klimchitskaya, and V.M. Mostepanenko, Phys. Lett. \textbf{A200}, 95 (1995) 

\bibitem{KlimPRA99} G.M. Klimchitskaya, A. Roy, U. Mohideen, and  V.M. Mostepanenko,
Phys. Rev. \textbf{A60}, 3487 (1999)  

\bibitem{MaiaNetoEPL05} P.A. Maia Neto, A. Lambrecht, and S. Reynaud, Europhys. Lett. \textbf{69}, 924 (2005) 

\bibitem{Golestanian} R. Golestanian and M. Kardar, Phys. Rev. Lett. \textbf{78}, 3421 (1997); 
Phys. Rev. \textbf{A58}, 1713 (1998) 


\bibitem{ChenPRL02} F. Chen, and U. Mohideen, Phys. Rev. Lett. \textbf{88}, 101801 
(2002) 

\bibitem{EmigPRA03} T. Emig, A. Hanke, R. Golestanian, and M. Kardar, Phys. Rev. 
\textbf{A67}, 022114 (2003) 

\bibitem{EmigPRL05} R. B\"uscher and T. Emig, 
Phys. Rev. Lett. \textbf{94}, 133901 (2005) 




\bibitem{MundayPRA2005} J.N. Munday, D. Iannuzzi, and F. Capasso, Phys. Rev. \textbf{A71}, 042102 (2005) 

\bibitem{RodriguezPRL2007} A. Rodriguez, M. Ibanescu, D. Iannuzzi, F. Capasso, J.D. Joannopoulos, and S.G. Johnson, Phys. Rev. Lett. \textbf{99}, 080401 (2007) 

\bibitem{DalvitPRL08} D.A.R. Dalvit, P.A. Maia Neto, A. Lambrecht, and S. Reynaud, Phys. Rev. Lett. \textbf{100}, 040405 (2008) 

\end{thebibliography}
\end{document}